\documentclass[latin1]{aa}
\usepackage{txfonts}

\usepackage[dvips]{epsfig}
\usepackage{longtable}
\usepackage{rotating}
\usepackage{soul}
\usepackage{graphicx}
\usepackage{color}
\usepackage{multicol}
\usepackage{lipsum}
\usepackage{dcolumn}
\usepackage{textcomp}

\usepackage{multirow}
\usepackage{amsmath}

\titlerunning{C-cyanomethanimine: laboratory and astronomical measurements}

\authorrunning{Melosso et al.}

\begin{document}

\newcommand{\icm}[1]{#1~\ensuremath{\mathrm{cm}^{-1}}}
\definecolor{lightblue}{rgb}{0.659,0.8706,1}
\definecolor{blu}{rgb}{0.2039,0.388,1}
\sethlcolor{lightblue} \setstcolor{red}
\newcommand{\toadd}[1]{\hl{#1}}
\newcommand{\todel}[1]{{\color{red}\st{#1}}}
\newcommand{\upd}[2]{{\color{red}\st{#1}}{\hl{#2}}}
\newcommand{\nota}[1]{\textit{\color{red} #1}}


\title{Laboratory measurements and astronomical search for cyanomethanimine
\thanks{The complete list of the measured transitions for $Z$- and $E$-C-cyanomethanimine is only available in electronic form
at the CDS via anonymous ftp to cdsarc.u-strasbg.fr (130.79.128.5) or via http://cdsweb.u-strasbg.fr/cgi-bin/qcat?J/A+A/}}


\author{M. Melosso
          \inst{1}
          \and
        A. Melli
          \inst{1}
          \and
        C. Puzzarini
          \inst{1,2,*}
          \and
        C. Codella
          \inst{2}
          \and
        L. Spada
          \inst{1,3}
          \and
        L. Dore
          \inst{1}
          \and
        C. Degli Esposti
          \inst{1}
          \and
        B. Lefloch 
          \inst{4}
          \and
        R. Bachiller 
          \inst{5}
          \and
        C. Ceccarelli
          \inst{4,2}
          \and
        J. Cernicharo
          \inst{6}
          \and
        V. Barone
          \inst{3}
        }

\institute{Dipartimento di Chimica ``Giacomo Ciamician", Universit\`{a} di Bologna, Via Selmi 2, I-40126 Bologna, Italy \\
           \email{cristina.puzzarini@unibo.it}  
         \and
           INAF, Osservatorio Astonomico di Arcetri, Largo E. Fermi 5, 50125, Firenze, Italy 
         \and
           Scuola Normale Superiore, Piazza dei Cavalieri 7, I-56126 Pisa, Italy 
         \and
         Univ. Grenoble Alpes, CNRS, Institut de
         Plan\'etologie et d'Astrophysique de Grenoble (IPAG), 38000 Grenoble, France 
         \and
         IGN, Observatorio Astron\'omico Nacional, Calle Alfonso XII, 28004 Madrid, Spain 
         \and
           Grupo de Astrof\'isica Molecular. Instituto de CC. de Materiales de Madrid (ICMM-CSIC). Sor Juana In\'es de la Cruz 3, Cantoblanco, 28049 Madrid, Spain
           }



\abstract
{C-cyanomethanimine (HNCHCN), existing in the two $Z$ and $E$ isomeric forms, is a key prebiotic molecule, but,
so far, only the $E$ isomer has been detected toward the massive star-forming region. Sagittarius B2(N) using transitions in the radio wavelength domain.} 
{With the aim of detecting HNCHCN in Sun-like-star forming regions, the laboratory investigation of its rotational spectrum
has been extended to the millimeter-/submillimeter-wave (mm-/submm-) spectral window in which several unbiased spectral surveys have been already carried out.}
{High-resolution laboratory measurements of the rotational spectrum of C-cyanomethanimine were carried out in the 100-420 GHz range using a frequency-modulation absorption spectrometer. We then searched for the C-cyanomethanimine spectral features in the mm-wave range using the
high-sensitivity and unbiased spectral surveys obtained with the IRAM 30-m antenna in the ASAI context, the earliest stages of star
formation from starless to evolved Class I objects being sampled.}
{For both the $Z$ and $E$ isomers, the spectroscopic work has led to an improved and extended knowledge of the spectroscopic parameters, 
thus providing accurate predictions
of the rotational signatures up to $\sim$700 GHz. So far, no C-cyanomethanimine emission
has been detected toward the ASAI targets, and upper limits of the column density
of $\sim$ 10$^{11}$--10$^{12}$  cm$^{-2}$ could only be derived. Consequently, the C-cyanomethanimine abundances have to be
less than a few 10$^{-10}$ for starless and hot-corinos. A less stringent constraint, $\leq$ 10$^{-9}$, 
is obtained for shocks sites.}
{The combination of the
upper limits of the abundances of C-cyanomethanimine together with accurate
laboratory frequencies up to $\sim$ 700 GHz poses the basis for future
higher sensitivity searches around Sun-like-star forming regions. 
For compact (typically less than 1$\arcsec$) and chemically enriched sources 
such as hot-corinos, the use of interferometers as NOEMA and ALMA in their 
extended configurations are clearly needed.}





\titlerunning{C-cyanomethanimine: laboratory and astronomical measurements}

\authorrunning{Melosso et al.}

\maketitle

\renewcommand{\textfraction}{0.15}
\renewcommand{\topfraction}{0.85}
\renewcommand{\bottomfraction}{0.65}
\renewcommand{\floatpagefraction}{0.60}

\section{Introduction}

Among the goals of astrochemistry, the detection of potential prebiotic molecules 
in astrophysical environments, and in particular in star forming regions, is fundamental  
in view of possibly understanding the origin of life. In recent years, several large programs
have been devoted to the detection of prebiotic species: 
Prebiotic Interstellar MOlecular Survey (PRIMOS\footnote{http://www.cv.nrao.edu/PRIMOS/}) project 
with the NRAO Green Bank Telescope (GBT), 
The IRAS16293-2422 Millimeter And Submillimeter Spectral 
Survey\footnote{http://www-laog.obs.ujf-grenoble.fr/heberges/timasss} (TIMASSS) with the IRAM 30-m and JCMT
single-dishes,  
CHESS\footnote{http://www-laog.obs.ujf-grenoble.fr/heberges/chess} 
(The Herschel Chemical Surveys of Star forming regions), 
ASAI\footnote{http://www.oan.es/asai/}
(Astronomical Surveys At IRAM) with the IRAM 30-m antenna, and, more recently,
SOLIS\footnote{http://solis.osug.fr/} (Seeds Of Life In Space) with IRAM NOEMA
(NOrthern Extended Millimeter Array), and PILS\footnote{http://youngstars.nbi.dk/PILS} 
(The ALMA Protostellar Interferometric Line Survey). 
These projects contributed to the census of a large number of new interstellar molecules (containing between six and eleven atoms),
see for example the Cologne Database for Molecular Spectroscopy \citep[CDMS,][]{2005JMoSt.742..215M}.

Among the various chemical species, the compounds containing the CN moiety are  
considered prebiotic molecules as potential precursors of amino acids
\citep[see, for example,][ and references therein]{balucani2009}.
The simplest one is HCN, which is ubiquitous in the interstellar medium (ISM). 
A particular case is represented by cyanomethanimine.
Among the HCN dimers, the $Z$- and $E$-C-cyanomethanimine forms (HNCHCN) as well as
N-cyanomethanimine (CH$_2$NCN) are isomers more stable than two isolated HCN molecules \citep{evans1991hcn}.
On general grounds, hydrogen cyanide dimers are thought to play a
role as intermediates in the prebiotic synthesis of purines and proteins \citep{ferris1984hcn}.
Within this context, cyanomethanimine, and in particular its C-form (C-HNCHCN) can be considered unique in the
family of COMs, and its detection around Sun-like protostars would be crucial in understanding
the prebiotic chemistry in regions that will form planetary systems.

C-cyanomethanimine has been detected by \citet{zaleski2013detection}
toward the massive star-forming region. Sagittarius (Sgr) B2(N), placed at 8.5 kpc from the Sun, within the PRIMOS context. 
To our knowledge, that reported by \citet{zaleski2013detection} is the first and so 
far unique detection of interstellar HNCHCN in our Galaxy.
The authors observed emission due to low-excitation ($E_{\rm up}$ up to 7 K)
transitions of the $E$ isomer in the $\simeq$ 9.5--48 GHz spectral range.

The detection of HNCHCN toward Sgr B2(N) in the centimeter-wave spectral window calls for  
further searches at higher frequencies, in the mm-/submm-wave spectral range, in regions forming future Sun-like stars 
using both single-dishes, like the IRAM 30-m antenna, and interferometers,
such as the IRAM NOEMA 
and Atacama Large Millimeter/submillimeter Array (ALMA). 
In addition, taking into account the increased sensitivities 
and new spectral windows offered by these telescopes (ALMA can touch the THz region),
it might be interesting to extend the observation of rotational 
features at far higher frequencies than those of \citet{zaleski2013detection}. 
Laboratory studies for C-cyanomethanimine were indeed limited to the portion of rotational spectra below 100 GHz \citep{takeo1986microwave,takano1990microwave,zaleski2013detection}. Because extrapolations from low-frequency laboratory measurements might provide inaccurate higher frequencies, the extension of the experimental investigation of rotational spectra of $Z$- and $E$-C-cyanomethanimine is therefore warranted. To guide this extension to higher frequency, a preliminary computational investigation of the spectroscopic parameters was carried out \citep{puzzarini2015isomerism}, thus pointing out the effect of the centrifugal distortion terms as well as the limited reliability of the $A$ rotational constant for both isomeric species. 

Furthermore, rotational spectra of both isomers show a maximum of intensity at frequencies higher than 100 GHz. As it will be shown later in the manuscript, the $E$ isomer shows strong $a$- and $b$-type spectra \citep[$\mu_a$ = 3.25(5) D, $\mu_b$ = 2.51(2) D;][]{takano1990microwave} even at low temperatures, whose maxima shift from $\sim$120 and $\sim$450 GHz at T = 10 K to $\sim$410 GHz and $>$1 THz at T = 300 K, respectively. $Z$-C-cyanomethanimine presents a weak $b$-type spectrum \citep[$\mu_b$=0.4(5) D;][]{takano1990microwave}, while more intense (but still weaker than the {\it E} one by about one order of magnitude) is the $a$-type one \citep[$\mu_a$=1.35(10) D;][]{takano1990microwave}, whose maximum shifts from $\sim$100 GHz at T = 10 K to $\sim$430 GHz at T = 300 K. 

The sensitivity reached in absorption measurements against the strong
continuum source SgrB2 by \citet{zaleski2013detection} cannot be obtained toward dark clouds because of
the strong continuum emission itself, which is more than a factor of ten larger than the kinetic temperature of quiescent
clouds. However, mm line emissions are the best tracers for detecting HNCHCN,
or to provide significant upper limits to its abundance, in dark clouds.
In order to have accurate frequencies in the mm domain, we have performed
a new set of measurements in our laboratory (in the 100-420 GHz frequency range) for the $E$ and $Z$
isomers of HNCHCN, which we present in this work, thus improving and enlarging the existing dataset of spectroscopic parameters. 

Based on the spectroscopic results of this work, which allow us to provide 
accurate frequency predictions up to 700 GHz, we have carried out a 
search for HNCHCN emission toward a sample of 8 nearby
(distances less than 250 pc) Sun-like-star forming
regions in the earliest phases: from starless to more evolved Class 0
and I objects passing through Barnard 1, considered an hydrostatic core
in a stage before the protostellar one (see Sect. 3).  
To this purpose, we used the ASAI unbiased high-sensitivity 
spectral surveys at mm-wavelenghts.
In summary, the main goal of the manuscript is twofold:
(i) to improve the predictions for rotational transitions reaching
700 GHz for both the $Z$ and $E$ isomers, and (ii) to search for
HNCHCN (using the new frequencies) for the first time 
in a large sample of low-mass star-forming regions.

\section{Experiment}
 
\subsection{Production of C- cyanomethanimine}

C-cyanomethanimine is an unstable molecule that, in the present work, was produced by pyrolysis of dimethylcyanamide, (CH$_3$)$_2$NCN, as described by \citet{takeo1986microwave}, by flowing the vapors of the precursor through a quartz tube heated by a 30 cm long tube furnace. The apparatus is the same used to produce other molecules of astrophysical interest, such as methanimine (CH$_2$NH) \citep{dore2012accurate} and ketenimine (CH$_2$CNH) \citep{degli2014accurate}. The quartz reactor was connected to the usual gas inlet of the free-space absorption cell of the spectrometer and the pyrolysis products were pumped out continuously, but slowly, in order to provide their continuous flow  inside the 3.25 m long, 10 cm diameter, glass cell. 

For each isomer, the best working conditions were obtained by monitoring the absorption signal of a previously reported transition below 100 GHz \citep[from][]{takano1990microwave}.
Slightly different optimal conditions were employed for two isomers: highest yields of $Z$-C-cyanomethanimine were obtained by setting the furnace temperature to 1100 
\textdegree C and by flowing the precursor at a pressure of 60 mTorr through the quartz reactor, which corresponds to a pressure of 10 mTorr in the absorption cell. On the other hand, $E$-C-cyanomethanimine was found to have a higher production rate by using a lower pressure (20 mTorr in the quartz reactor) and a higher pyrolysis temperature (1160 \textdegree C). 

\subsection{Millimeter/submillimeter-wave spectrometer}

The rotational spectra were recorded in the 100-117 GHz and 240-419 GHz frequency regions by means of a millimeter/submillimeter-wave frequency modulated spectrometer \citep{degli2017millimeter}. Radiations sources are either a series of Gunn diodes covering the 75-134 GHz range or passive multipliers driven by the Gunn diodes which extend the covered range from 225 GHz to 1.2 THz. The output frequency is stabilized by a phase-lock loop (PLL) system referred to a signal of 75 MHz and frequency modulated at 6 kHz. Phase sensitive detection at twice the modulation frequency is employed, so that the second derivative of the actual absorption profile is recorded. A Schottky barrier Millitech detector was used for recording below 117 GHz, while VDI detectors were employed in the 240-419 GHz region.

\begin{figure*}[tbp]
\centering
\includegraphics[width=14cm]{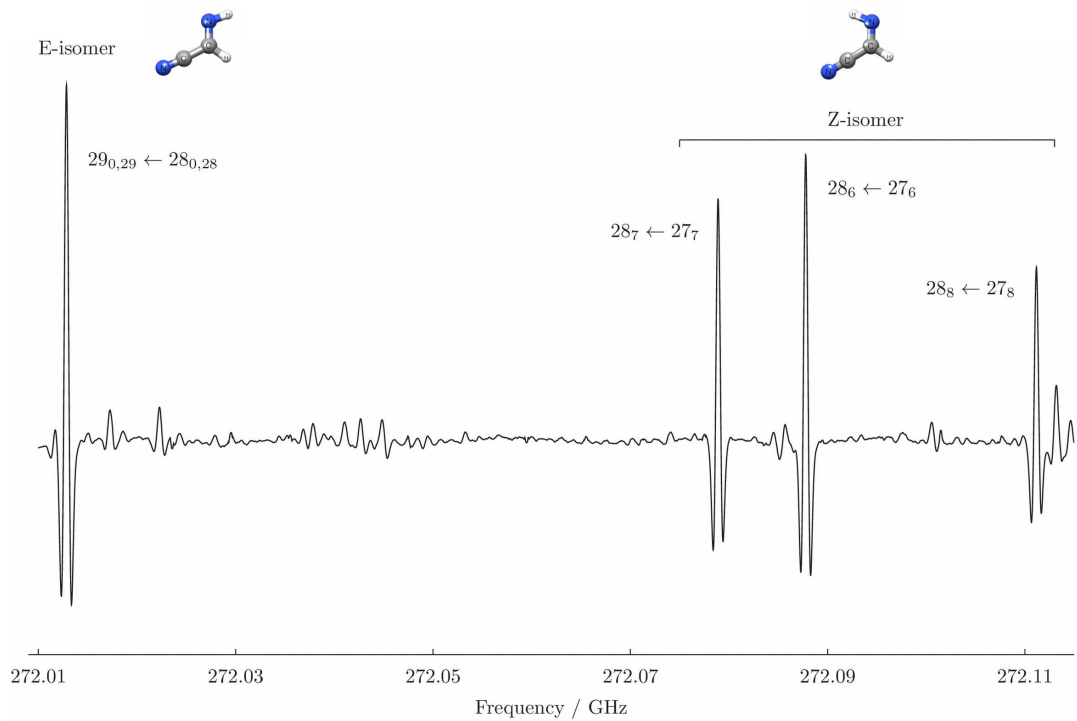}
\caption{Portion of the millimeter-wave spectrum, recorded in the best condition for the production of {\it E}- cyanomethanimine. {\it a}-type {\it R} branch transitions are visible for both isomers.} \label{figCOMP}
\end{figure*}

According to the experimental conditions, frequency range, and signal-to-noise ratio (S/N), the estimated uncertainties for our measurements range from 20 to 60 kHz. Figure~\ref{figCOMP} shows a small portion of spectra at 272 GHz. As seen, both isomers are present in the experimental mixture, with an abundance that ensures a very good S/N of the spectrum. Despite the fact that the recording was carried out under the best conditions for producing the $E$ isomer and the lower dipole moment of $Z$-C-cyanomethanimine, the transitions appear to have similar intensity, indicating that the $Z$ isomer is produced in higher yield.

\section{The astronomical sample}

The data presented here are part of the 
Large Programme ASAI, which has  
collected unbiased spectral surveys using
the Pico Veleta (Spain) 30-m single-dish. Observations
and data reduction are reported in details in 
Lefloch et al. (in preparation). In summary, the observations
were carried out in Wobbler Switching Mode, during several 
runs between 2012 and 2014 using
the EMIR receivers at 3mm (72--116 GHz), 2mm (126--173 GHz),
and 1.3mm (200--276 GHz).  
In the present context we searched for C- cyanomethanimine 
in different objects sampling different stages of the
formation process leading to a Sun-like star,
namely from starless cores to Class 0 and I objects,
and in addition to jet-driven shocks regions.
Below, a short description of the targets:

\begin{itemize}

\item
{\bf L1544}
is one of the best studied starless core located
in Taurus at a distance of 140 pc \citep[see][ and references within]{caselli1999co,caselli2002molecular} . 
The core is characterized by high molecular depletion given its
high density ($\geq$ 2 $\times$ 10$^6$ cm$^{-3}$) and
low temperature (down to 7 K). Recently, emission due to the so-defined interstellar complex organic molecules (iCOMs) 
(e.g., CH$_3$CHO, CH$_3$OCHO, CH$_3$OCH$_3$) have been detected, 
plausibly coming from
the external portion of the clouds due to reactive desorption
and/or cosmic-rays irradiation \citep[e.g.,][ and references within]{vastel2014origin,jimenez2016spatial,vasyunin2017formation}. 

\item
{\bf Barnard B1b}
is an active star-forming site \citep[e.g.,][]{bachiller1986relation}
located in the Barnard dense core, in Perseus 
($d$ = 235 pc). B1b is associated with two companions, B1bN and B1bS:
their spectral energy distribution \citep{pezzuto2012herschel} 
and their association with compact
and slow outflows \citep{gerin2015nascent} make them among the 
best candidates for the first
hydrostatic stage. In other words, Barnard B1b could be placed in
an intermediate stage between starless cores and Class 0 ($\geq$ 10$^4$ yr)
protostars. The observed position is associated with a rich molecular
spectra containing for example, CH$_3$CHO, CH$_3$OCHO, and 
CH$_3$O lines \citep[e.g.,][]{cernicharo2012discovery,daniel2013nitrogen}.

\item
{\bf IRAS4A}
is a binary Class 0 system in the Perseus NGC1333 region, 
well identified using
interferometry \citep[e.g.,][ and references therein]{looney2000unveiling,santangelo2015jet,tobin2016vla}.
The two objects, called A1 and A2, are separated by 1.8$''$ (420 AU) 
and are associated with different properties: 
(i) IRAS4-A1 has an internal luminosity 
of $\sim$ 3 $L_{\rm \sun}$ \citep{de2017glycolaldehyde} and
is more than three times brighter in the mm-flux than
its companion; (ii) only IRAS4-A2 is associated with a  
emission due to iCOMs \citep[e.g., HCOCH$_2$OH among others; 
see][]{taquet2015constraining,coutens2015detection,de2017glycolaldehyde}. 
Both sources drive
 jets, with A1 definitely being the faster and younger of the two \citep{santangelo2015jet}.  
The intrinsic different properties of jets and driving sources in  
NGC1333-IRAS4A indicates different evolutionary stages, with A2
being evolved enough to develop a hot-corino region.

\item
{\bf L1157-mm}
is a Class 0 source with a bolometric luminosity
of $\sim$ 3 $L_{\rm \sun}$ driving a precessing jet \citep{gueth1996precessing,gueth1998sio,podio2016first}, which, in turn, has created the prototype 
of the so-called chemically rich outflows \citep[e.g.,][]{bachiller2001chemically}.
The protostar lies in a relatively isolated cloud in Cepheus, at a
distance of 250 pc \citep{looney2000unveiling}, and is associated with an
elongated molecular envelope possibly associated with a still not detected
accretion disk \citep[e.g.,][ and references therein]{gueth2003dust,chiang2012apj,tobin2013resolved,tobin2013vla}.

\item
{\bf L1157-B1}
is the brightest shocked region associated with the jet/outflow
driven by the L1157-mm protostar. 
The jet has excavated two main cavities, with apices
called B1 and B2 \citep{gueth1996precessing,gueth1998sio}. 
In particular, B1 has a kinematical age
$\simeq$ 1100 yr \citep{podio2016first} and
consists of a series of shocks caused by different episodes of ejection
impacting against the cavity wall.
L1157-B1 has been the target
of several studies using single-dish and interferometric arrays
revealing rich and intense molecular spectra 
\citep[e.g.,][]{tafalla1995ammonia,bachiller2001chemically,codella2010chess,lefloch2010chess,nisini2010water}.
Interestingly, high-angular resolution images revealed 
a chemical differentation
indicating an active grain-surface chemistry at work
\citep[e.g.,][]{codella2009methyl,benedettini2007clumpy,benedettini2012chess,benedettini2013b1,busquet2014chess}.
Several iCOMs have been revealed, from
the first detections reported by \citet{arce2008complex} to the \citet{lefloch2017l1157}
extensive survey, passing through the 
first detection of formamide in a shock \citep{mendoza2014molecules}.

\item
{\bf L1448 R2}
is a shocked region located in the southern molecular
outflow driven by a Class 0 protostar with a luminosity of
about 7 $L_{\rm \sun}$ \citep[e.g.,][]{bachiller1990high,de2017glycolaldehyde}.
The region is located in Perseus, at the center of the
L1448 complex \citep[see][ and references therein]{looney2000unveiling,tobin2016vla}
 at a distance of 232 pc \citep{hirota2011astrometry}.  
L1448 R2 has been studied in detail down to high-spatial resolutions
revealing high-velocity bullets, high-excitation conditions, and 
a consequently enriched chemistry
\citep[e.g., SiO, H$_2$O, NH$_3$, high-J CO lines:][]{dutrey1997successive,nisini2010water,nisini2013mapping,santangelo2012herschel,gomez2016diagnosing}.
To our knowledge, no evidence of emission due to iCOMs has been found so far.

\item
{\bf L1527}
is a Class 0 source located in Taurus ($d$ = 140 pc) with a
bolometric luminosity of approximately 3 $L_{\rm \sun}$ \citep[][ and
references therein]{tobin2013vla}.
The source is considered to be the prototypical 
warm carbon chain chemistry (WCCC) source, and
is associated with an almost edge-on envelope/disk system 
 \citep[see e.g.,][ and references therein]{sakai2010distributions,sakai2014change,sakai2014chemical,oya2015geometric}.
Recent ALMA images revealed
the rich chemistry (SO, CH$_3$OH) 
activated by the slow
shocks occurring at the centrifugal barrier of the infalling and rotating
envelope \citep{oya2015geometric,sakai2017vertical}. 

\item
{\bf SVS13-A}
is part of the system SVS13, located in the NGC1333 cloud in Perseus
at 235 pc from the Sun.
In the mm-spectral range the region is dominated by two protostars 
identified by interferometric
observations \citep{bachiller1998molecular,looney2000unveiling,chen2009iram,tobin2016vla}, 
called A and B, at 15$\arcsec$ from each other.
The internal luminosity of SVS13-A has been
estimated to be around 25 L$_\odot$ \citep{de2017glycolaldehyde}. 
SVS13-A is associated with: (i) an extended ($>$0.07 
pc) outflow, (ii) the HH7-11 chain \citep[][and references therein]{lefloch1998cores}, and (iii) 
a low L$_{submm}$/L$_{bol}$ ratio ($\sim$ 0.8 $\%$).
As a consequence, although SVS13-A is still deeply embedded 
in a large-scale envelope \citep[$\sim$ 6000 AU;][]{lefloch1998cores},
the protostar is considered a Class I
($\geq$ 10$^5$ yr) source
\citep[e.g.,][and references therein]{chen2009iram}.
Recently, the occurrence of a hot-corino around SVS13-A has been
revealed through HDO and HCOCH$_2$OH measurements \citep{codella2016hot,de2017glycolaldehyde}.

\end{itemize}

\section{Results}

\subsection{Spectroscopic parameters}
\begin{figure}[tbp]
\centering
\includegraphics[width=9cm]{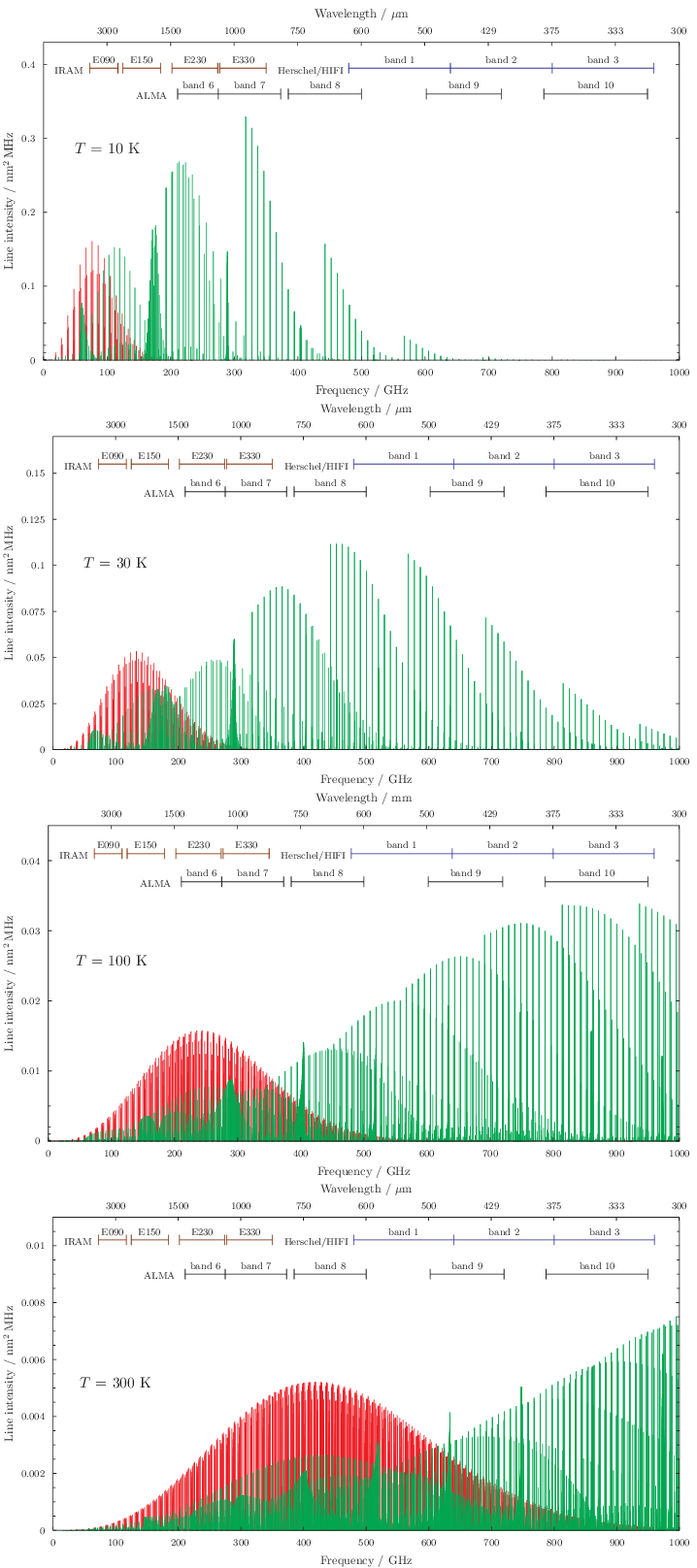}
\caption{Spectral predictions in the 0 - 1000 GHz frequency range for the {\it E}-C-cyanomethanimine at four different temperatures: T = 10 K (upper row), T = 30 K (second row), T = 100 K (third row), and T = 300 K (lower row). $a$-type transitions are depicted in red, $b$-type transitions in green.} \label{figPREV-E}
\end{figure}

A total number of 286 and 311 new line frequencies were measured in the 100-419 GHz range for the $E$ and $Z$ isomers, respectively.
Our measurements involved rotational energy levels in the 10 to 46 $J$ interval
and ranging in $K_a$ from 0 to 15. Small splittings due to the
$^{14}$N-nuclear spin of the two nitrogen atoms were observed for a
few $a$- and $b$-type $R$ branch transitions. 
The transition frequencies obtained in the present study together with those reported by \citet{takeo1986microwave} in the 37-50 GHz
range, by \citet{takano1990microwave} in the 23-100 GHz range, and by \citet{zaleski2013detection} in the 9-48 GHz range were analyzed
using Pickett's SPFIT program \citep{pickett1991fitting}, adopting Watson's S-reduced Hamiltonian in its I$^r$ representation
\citep{watson1977aspects}.
Each transition frequency was weighted proportionally to the inverse square of its experimental uncertainty. 
The hyperfine structure exhibited by many lines of both isomers was accounted for using the following angular momenta coupling scheme between the $^{14}$N-nuclear spins $\hat{I_1}$ (NH group) and $\hat{I_2}$ (CN
group) and the rotational angular momentum $\hat{J}$: 
$\hat{I}$ = $\hat{I_1}$+$\hat{I_2}$, $\hat{F}$ = $\hat{J}$+$\hat{I}$.
In this way, a total of 384 distinct transition frequencies were analyzed for each of the two isomers.
These global fits led to the determination of 17 independent spectroscopic parameters with a root mean 
square (RMS) error of 36 and 38 kHz and a dimensionless standard deviation of 0.81 and 0.95 for $Z$- and $E$-C-cyanomethanimine, respectively. The results of the fits, compared to those 
by \citet{zaleski2013detection}, are collected in Table~\ref{t1}, while the list of frequencies is available in the supplementary material, which also contains the set of spectroscopic constants obtained in
the SPFIT format (in order to facilitate their inclusion in spectroscopic databases). We note that, in addition to the improvement of the already known spectroscopic parameters, new constants have been determined. In particular, the quartic centrifugal distortion constant $D_K$ and four sextic centrifugal-distortion constants were obtained with good accuracy for both isomers. The $^{14}$N spin-rotation coupling constant $C_{aa}$ (NH) was also fitted.
Overall, the accuracy of all spectroscopic constants previously reported has been improved by 1 to 3 order of magnitude. For both isomers, there is a noticeable difference in the newly determined $A$ rotational constant with respect to those of \citet{zaleski2013detection}, with differences of $\sim$5 MHz for $E$-C-cyanomethanimine and $\sim$20 MHz for the $Z$ isomer, the respective discrepancies being about 200 and 4 times the statistical errors given in \citet{zaleski2013detection}.
These large changes are due to the fact that $b$-type transitions have been measured for the first time for $Z$-C-cyanomethanimine, while  $b$-type transitions involving rotational levels with $K_a$ > 1 have been included in the fit for the $E$ isomer.   
Our global fits allowed us to improve the rest frequency determinations in terms of accuracy as well as to extend their availability up to 700 GHz with uncertainty smaller than 200 kHz.  
By means of a selection of observed transition frequencies together with the corresponding residuals from the fits, Table~\ref{t2} provides the reader with an example of the quality of the measurements and analysis for both isomeric species.

\begin{figure}[tbp]
\centering
\includegraphics[width=9cm]{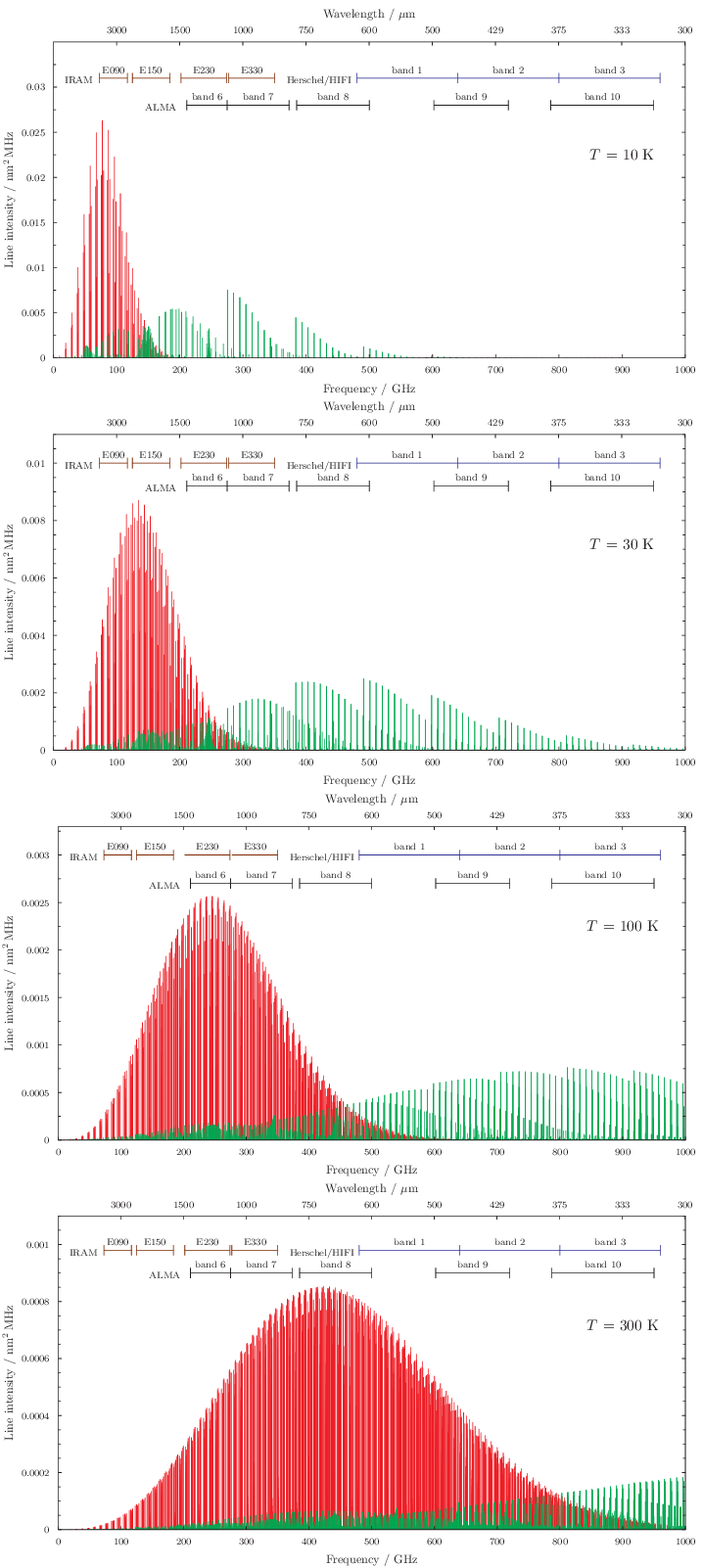}
\caption{Spectral predictions in the 0 - 1000 GHz frequency range for the {\it Z}-C-cyanomethanimine at four different temperatures: T = 10 K (upper row), T = 30 K (second row), T = 100 K (third row), and T = 300 K (lower row). $a$-type transitions are depicted in red, $b$-type transitions in green.} \label{figPREV-Z}
\end{figure}

\begin{table*}[t]
        \caption{\label{t1}Spectroscopic parameters determined for the Z and E isomers of C-cyanomethanimine.}
        \centering
        \begin{tabular}{p{1.5cm}l|D{.}{.}{10}D{.}{.}{10}D{.}{.}{10}D{.}{.}{10}}
                \hline\hline
                \noalign{\smallskip}
                \multicolumn{2}{l}{} & \multicolumn{2}{c}{E-C-cyanomethanimine} & \multicolumn{2}{c}{Z-C-cyanomethanimine} \\
                \hline
                \noalign{\smallskip}
                \multicolumn{2}{l}{Parameters} & \multicolumn{1}{c}{Present work} & \multicolumn{1}{c}{Previous results\tablefootmark{a}} & \multicolumn{1}{c}{Present work} & \multicolumn{1}{c}{Previous results\tablefootmark{a}} \\
                \hline
                \noalign{\smallskip}
                $         A   $ &                 &    62700.392(22)\tablefootmark{b}               &  62695.094(24)     &     54193.405(32)             &   54173.1(50)  \\
                $         B   $ &                 &   4972.04534(22)               & 4972.04643(81)     &   5073.86584(15)             & 5073.86506(86) \\
                $         C   $ &                 &   4600.29561(23)               & 4600.29460(89)     &   4632.38905(14)             & 4632.39090(74) \\
                $       D_{J} $ & $\times 10^{3}$ &        1.881477(94)             &      1.8704(55)    &      2.425671(87)             &   2.4737(74)  \\
                $       D_{JK}$ &                 &      -0.1054288(30)            &    -0.10455(17)    &   -0.1032838(27)             &  -0.10331(21)   \\
                $       D_{K} $ &                 &          5.1408(52)             &                    &        3.5488(59)             &                \\
                $       d_{1} $ & $\times 10^{3}$ &       -0.33939(13)             &     -0.3272(57)    &     -0.48265(10)             &   -0.4961(82)  \\
                $       d_{2} $ & $\times 10^{3}$ &      -0.020511(20)             &      0.0221(71)    &    -0.031338(21)             &  -0.0321(55)   \\
                $       H_{J} $ & $\times 10^{9}$ &           4.661(33)             &                    &         6.998(35)             &                \\
                $      H_{JK} $ & $\times 10^{6}$ &        -0.42172(97)             &                    &      -0.40281(93)             &                \\
                $      H_{KJ} $ & $\times 10^{6}$ &         -7.303(11)             &                    &       -8.986(15)             &                \\
                $       H_{K} $ & $\times 10^{3}$ &        0.906723\tablefootmark{c} &                    &      0.599872\tablefootmark{c} &                \\
                $       h_{1} $ & $\times 10^{9}$ &           1.823(45)             &                    &         2.555(38)             &                \\
                $       h_{2} $ & $\times 10^{9}$ &          0.1571\tablefootmark{c} &                    &        0.2675\tablefootmark{c} &                \\
                $       h_{3} $ & $\times 10^{9}$ &          0.0450\tablefootmark{c} &                    &        0.0738\tablefootmark{c} &                \\
                $\chi_{aa}(CN)$ &                 &         -4.1315(20)            &    -4.1280(67)     &       -4.0102(51)             &   -4.012(21)             \\
                $\chi_{bb}(CN)$ &                 &          -0.2006(31)              &    -0.1972(57)     &         -0.20845(39)            &  -0.2146(86)              \\
                $\chi_{aa}(NH)$ &                 &          0.7447(30)            &      0.753(11)    &       -4.2721(49)             &    -4.269(21)            \\
                $\chi_{bb}(NH)$ &                 &          -2.0661(51)              &      -2.0642(89)   &        -0.81008(50)             &   -0.8201(82)             \\
                $  C_{aa}(NH) $ &                 &            0.0173(19)            &                    &         0.0064(22)             &                \\
                \hline
                \noalign{\smallskip}
                \# lines               &                 &  \multicolumn{1}{c}{384}          &                 &   \multicolumn{1}{c}{384}                  &                \\
                RMS error          &                 &  \multicolumn{1}{c}{0.036}                &                    &           \multicolumn{1}{c}{0.038}                &                \\
                $\sigma$            &                 &  \multicolumn{1}{c}{0.81}               &                    &   \multicolumn{1}{c}{0.95}                &                \\
                \hline
        \end{tabular}
        \tablefoot{Units are in MHz, except the dimensionless standard deviation $\sigma$. \\
                \tablefoottext{a}{\citet{zaleski2013detection}}
                \tablefoottext{b}{Values in parenthesis denote one standard deviation and apply to the last digits of the constants.}
                \tablefoottext{c}{Fixed at the computed value \citep{puzzarini2015isomerism}}
        }
\end{table*}

\begin{table*}[t]
        \caption{\label{t2} Selection of observed transition frequencies together with the corresponding residuals from the fits for the $Z$ and $E$ isomers of C-cyanomethanimine.}
        \centering
        \scalebox{0.9}{
        \begin{tabular}{cccccccccc}
                \hline\hline
                \noalign{\smallskip}
                Isomer & J$^{\prime}$ & K$^{\prime}_a$ & K$^{\prime}_c$ & J & K$_a$ & K$_c$ & \multicolumn{1}{c}{Obs. frequency} & \multicolumn{1}{c}{Uncertainty} & \multicolumn{1}{c}{Obs.-Calc.} \\
                \noalign{\smallskip}
                \multicolumn{1}{c}{} & \multicolumn{3}{l}{Upper state} & \multicolumn{3}{l}{Lower state} & \multicolumn{1}{c}{(MHz)} & \multicolumn{1}{c}{(MHz)} & \multicolumn{1}{c}{(MHz)}  \\
                \hline
                \noalign{\smallskip}
                ($Z$) & 30 &  3 & 28 &  29 &  3 & 27 &    291623.715  &   0.030  &   \phantom{-}0.008  \\
                ($Z$) & 30 &  5 & 26 &  29 &  5 & 25 &    291640.038  &   0.030  &  -0.018   \\
                ($Z$) & 30 &  5 & 25 &  29 &  5 & 24 &    291646.827  &   0.030  &  -0.007   \\
                ($Z$) & 30 &  4 & 27 &  29 &  4 & 26 &    291842.010  &   0.030  &  -0.012   \\
                ($Z$) & 30 &  4 & 26 &  29 &  4 & 25 &    292014.364  &   0.030  &  -0.051   \\
                ($E$) & 42 &  2 & 41 &  42 &  1 & 42 &    345022.426  &   0.030  &  -0.006   \\
                ($E$) & 36 &  5 & 31 &  35 &  5 & 30 &    345027.550  &   0.040  &   \phantom{-}0.008   \\
                ($E$) & 36 &  9 & 27 &  35 &  9 & 26 &    345039.468  &   0.040  &   \phantom{----}0.000\tablefootmark{(a)}  \\
                ($E$) & 36 &  9 & 28 &  35 &  9 & 27 &    345039.468  &   0.040  &   \phantom{----}0.000\tablefootmark{(a)}  \\
                ($E$) & 37 &  0 & 37 &  36 &  0 & 36 &    345190.487  &   0.040  &   \phantom{-}0.006    \\              
                \hline\hline
        \end{tabular}
}
\tablefoot{     
\tablefoottext{a}{Blended transitions: relative weight 0.5.} 
}
\end{table*}

\begin{figure*}[tbp]
\centering
\includegraphics[width=14cm]{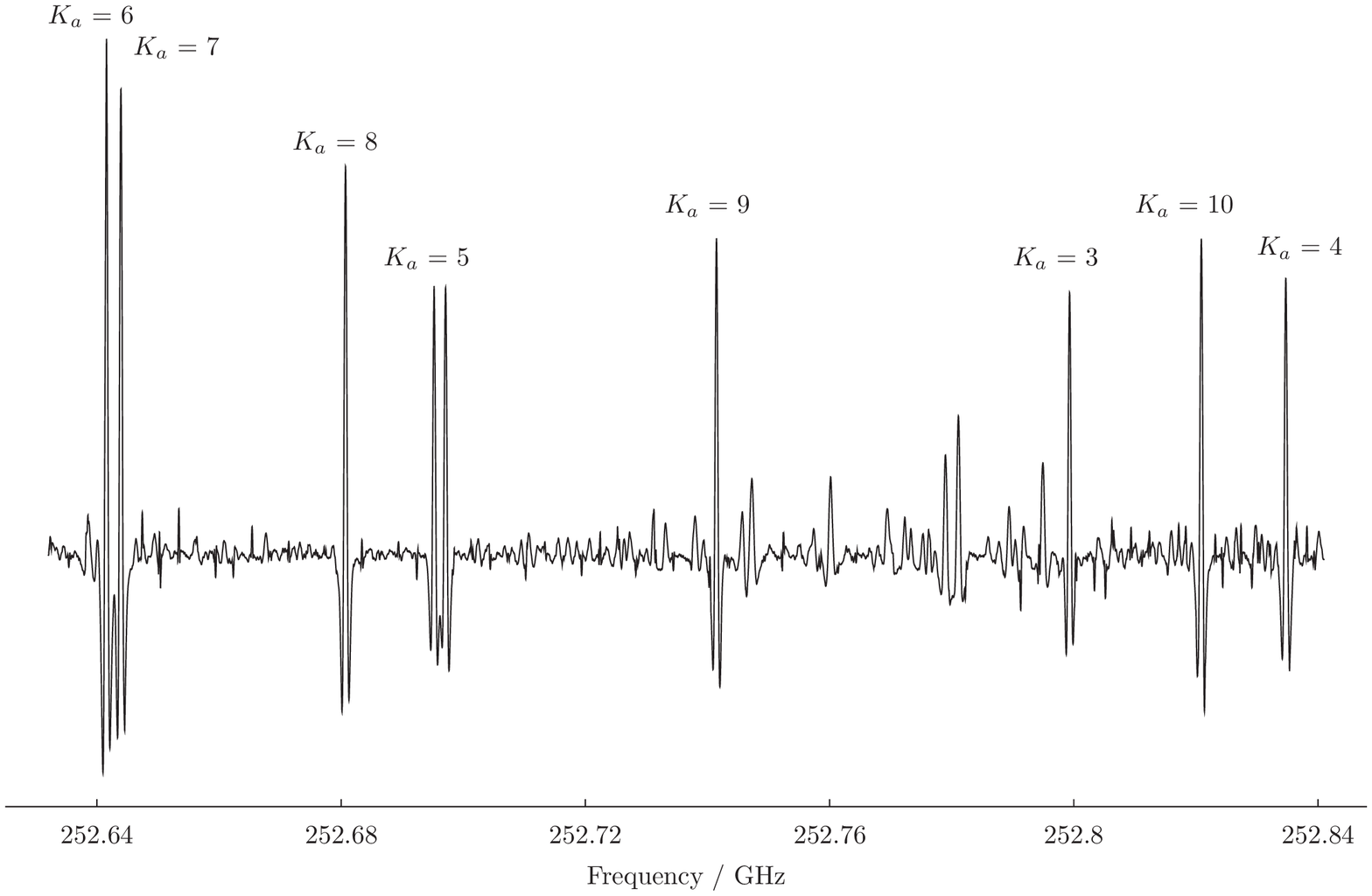}
\caption{Portion of the $J$=26$\gets$25 {\it a}-type band of {\it Z}-cyanomethanimine. Both components of the asymmetry doublet are visible for the $K_a$=5 transition, while only the low-frequency component is visible for $K_a$=3 and $K_a$=4 transitions.} \label{figLS}
\end{figure*}

Figures~\ref{figPREV-E} and~\ref{figPREV-Z} provide an overview of the rotational spectra in the 0-1000 GHz frequency range for the $E$ and $Z$ isomers, respectively, at four different temperatures: T = 10 K (upper row), T = 30 K (second row), T = 100 K (third row), and T = 300 K (lower row). For both isomers, it is observed that by increasing the temperature the maximum of intensity moves to higher frequencies. The overall spectra also reflect the dipole moment components of the two species. For $E$-C-cyanomethanimine, both $a$- and $b$-type spectra are intense (see Figure~\ref{figPREV-E}), as expected by the large dipole moment components. However, the $b$-type is the most intense at all temperatures considered, even if at T = 300 K the $a$-type transitions become nearly as intense as the $b$-type ones (see Figure~\ref{figPREV-E}, lower row). For the $Z$ isomer, the $a$-type spectrum is the more intense independently of the temperature because of the small magnitude of $\mu_b$ (see Figure~\ref{figPREV-Z}). Despite the fact that the $b$-type spectrum is particularly weak, in this work, for the first time, $b$-type transitions have been measured and the observed intensities tend to confirm qualitatively a value of 0.4 D for $\mu_b$. Overall, the rotational spectrum of $Z$-C-cyanomethanimine is weaker than that of the $E$ isomer by about one order of magnitude. An example is shown in Figure~\ref{figLS}, which allows us to point out the good S/N of our measurements.

\subsection{Astronomical observations: abundance upper limits}

To obtain the most constraining information on the presence of C-cyanomethanimine in the selected astronomical targets,
we used the frequencies (falling in the ASAI spectral ranges) of the $E$ isomer, whose transitions
are always brighter than those of the $Z$ one.
Based on the predictions showed of Figure~\ref{figPREV-E}, and conservatively homogenising the
search for HNCHCN in the present sample, we selected the brightest lines assuming a temperature 
(and the corresponding partition function) of 10 K for the starless core L1544 and for Barnard 1, 
while we used a representative temperature of 100 K for the regions
around protostars and outflow shocks. 

Figure~\ref{figPREV-E} clearly shows that the most intense lines at 10 K fall in
the 3mm spectral window (where the Half Power Beam Width, HPBW, is $\sim$ 26$\arcsec$), while 
at 100 K they lie at 1.3mm (HPBW $\sim$ 12$\arcsec$). 
As an example, the 9$_{\rm 0,9}$--8$_{\rm 0,8}$ transition at 85931.777 MHz with $E_{\rm u}$ = 115 K
and $S\mu$$^2$ = 233 D$^2$ is expected to be one of the brightest lines at 10 K.
On the other hand, a good candidate at 100 K is
the 22$_{\rm 0,22}$--21$_{\rm 0,21}$ transition at 207705.762 MHz
($E_{\rm u}$ = 21 K, $S\mu$$^2$ = 190 D$^2$). 
Unfortunately, no lines due to C-cyanomethanimine were detected.
However, thanks to the high-sensitivity of the ASAI dataset, we
derived constraining (see below) upper limits on the HNCHCN column
density in low-mass star-forming regions. 
To obtain the best sensitivity, we smoothed the spectral resolution of the L1544
spectra to 0.5 km s$^{-1}$, given the
expected linewidth of iCOMs lines is between 0.3 km s$^{-1}$
and 0.8 km s$^{-1}$ \citep{vastel2014origin}. 
On the other hand, we smoothed the spectra of 
Barnard 1, protostars and shocks to
2 km s$^{-1}$ because we expect a linewidth of at least 
4 km s$^{-1}$ \citep[e.g.,][]{de2017glycolaldehyde,lefloch2017l1157, 
cernicharo2012discovery}.
The 1$\sigma$ level of the integrated area (in main-brightness temperature scale, $T_{\rm MB}$) 
of the lines are: $\sim$ 10--20 mK km s$^{-1}$ for L1544 and
Barnard1, and $\simeq$ 30--90 mK km s$^{-1}$
for the other targets.
Using a 3$\sigma$ level criterium, we obtained
the following beam averaged (12$\arcsec$ for all the sources but L1544
and Barnard 1, averaged
on 26$\arcsec$) upper limits: 
\begin{itemize}
\item 1 $\times$ 10$^{11}$ cm$^{-2}$ (L1544; 26$\arcsec$)
\item 2 $\times$ 10$^{11}$ cm$^{-2}$ (Barnard 1; 26$\arcsec$)
\item 1 $\times$ 10$^{12}$ cm$^{-2}$ (L1157-B1; 12$\arcsec$)
\item 2 $\times$ 10$^{12}$ cm$^{-2}$ (L1157-mm, L1448-R2; 12$\arcsec$)
\item 3 $\times$ 10$^{12}$ cm$^{-2}$ (IRAS4A, L1527, SVS13-A; 12$\arcsec$)
\end{itemize}

An estimate of the upper limits for the cyanomethanimine abundance, $X_{\rm HNCHCN}$,
can be obtained using the H$_2$ column density: namely N(H$_2$) $\simeq$ 10$^{23}$ cm$^{-2}$ for the starless
core L1544 \citep{2005ApJ...619..379C}, $\simeq$ 10$^{21}$ cm$^{-2}$ for the L1157-B1 shock \citep{2012ApJ...757L..25L}, 
and around 10$^{24}$ cm$^{-2}$ for the hot-corinos
associated with SVS13-A and IRAS4A \citep[e.g.,][]{looney2000unveiling,2002A&A...395..573M,codella2010chess}.
For a proper comparison, the beam averaged $N_{\rm HNCHCN}$ upper limits have to be first modified
taking into account the expected emitting size, being consequently corrected for the 
corresponding beam dilution. For hot-corinos, we assumed a typical size of 1$\arcsec$ and then
the filling factor\footnote{ $ff$ = $\theta_{\rm s}$$^2$/($\theta_{\rm s}$$^2$+$\theta_{\rm b}$$^2$), where $\theta_{\rm s}$
and $\theta_{\rm b}$ are the source and the beam sizes, respectively.} 
$ff$ = 7 $\times$ 10$^{-3}$; for
the L1157-B1 shock, we used 9$\arcsec$ and $ff$ = 0.36 \citep{2012ApJ...757L..25L}. A filling
factor $ff$ = 1 (i.e., no correction) has been assumed instead for L1154 given its extended
structure \citep[e.g.,][]{vastel2014origin}.
Therefore, by assuming that HNCHCN and H$_2$ are tracing the same material and comparing the
corresponding column densities, 
we derived $X_{\rm HNCHCN}$ $\leq$ 4 $\times$ 10$^{-10}$ 
for starless and hot-corinos, 
and $\leq$ 5 $\times$ 10$^{-9}$ for shocks.

\subsection{Discussion}
As stated in the Introduction, the unique detection of  
cyanomethanimine in the interstellar medium so far has been reported by \citet{zaleski2013detection}, who observed emission due to the $E$ isomer 
toward the B2(N) core of the Sagittarius complex. 
Sagittarius B2(N) can be considered one of the best places where
to search for complex organic molecules \citep[e.g.,][]{2009A&A...499..215B}.
The Sagittarius B2 region is one of the largest molecular
clouds in the Galaxy associated with massive star-forming regions,
and it is located at about 120 pc from the
Galactic Center. \citet{zaleski2013detection} derived a $N_{\rm HNCHCN}$
$\simeq$ 10$^{13}$ cm$^{-2}$ using lines observed at 1cm using
the Green Bank Telescope (GBT), but no abundance has been 
calculated. This is plausibly due to the uncertainty on the size of the 
emitting region, Sgr B2(N) being a source with a substantial structure on
spatial scales smaller than the GBT beam 
(from $\simeq$ 20$\arcsec$
to 80$\arcsec$, depending on the frequency, i.e. 
$\simeq$ 1--3 pc given the Sgr B2(N) distance).
As a consequence, it is not clear which is the H$_2$ column density
of the observed features: if compact, $N_{\rm H_2}$ is surely
larger than 10$^{24}$ cm$^{-2}$, which in turn means
$X_{\rm HNCHCN}$ $\leq$ 10$^{-11}$. However, we cannot exclude
that the HNCHCN lines in Sgr B2(N) arise from the external layers
of the cloud where $N_{\rm H_2}$ could be lower. 

The advantages of the upper limits on $X_{\rm HNCHCN}$  presented here
are the following: (i) they refer to regions associated with Sun-like progenitors which are
expected to be associated with protoplanetary regions, and (ii) previous observations of the
astronomical sample make one confident to give a reasonable 
assumption on the cyanomethanimine emitting region. 

Interestingly, the present upper limits on $X_{\rm HNCHCN}$ can be
compared with the abundance of another
N-bearing iCOMs, such as formamide (NH$_2$CHO), measured by
\citet{mendoza2014molecules} and \citet{2015MNRAS.449.2438L}
toward shocks ($X_{\rm NH_2CHO}$ = 5 $\times$ 10$^{-9}$) and 
hot-corinos (3 $\times$ 10$^{-11}$ and 2--5 $\times$ 10$^{-10}$ 
for SVS13-A and IRAS4A, respectively).
As a consequence, we have $R$ = $X_{\rm HNCHCN}$/$X_{\rm NH_2CHO}$
$\leq$ 10 (SVS13-A), while for
L1157-B1 and IRAS4A we have one order of magnitude less: $R$ $\leq$ 1.
Although these measurements do not severely
constrain the ratio between cyanomethanimine and formamide 
around Sun-like-star forming regions, they are in agreement with 
what we can derive from the column densities reported 
toward the massive star-forming region Sgr B2(N) 
by \citet{zaleski2013detection} and \citet{2011ApJ...743...60H}.
With the same caveats reported above on the use of large beams
around Sgr B2(N): $N_{\rm HNCHCN}$ = 2 $\times$ 10$^{13}$ cm$^{-2}$,
$N_{\rm NH_2CHO}$ = 4 $\times$ 10$^{14}$ cm$^{-2}$, and thus $R$ $\sim$ 0.1.

\section{Conclusions}

In the present paper, the investigation of the rotational spectrum 
of C-cyanomethanimine has been extended to the millimeter-/submillimeter-wave frequency region, thus 
considering the 100-419 GHz range.
New measurements, which also include the first recording of $b$-type transitions for the $Z$ isomer, have allowed us to improve and enlarge the sets of spectroscopic parameters. Overall, the present work is able to provide accurate predictions for rotational
transitions up to 700 GHz for both $Z$- and $E$-C-cyanomethanimine. This is an important prebiotic species, whose $E$ form has already been detected 
in SgB2(N) at frequencies below 48 GHz. The extension to higher frequencies provided by this work opens up the opportunity of also detecting the $Z$ isomer,
which has an $a$-type rotational spectrum weaker than that of $E$-C-cyanomethanimine by a factor of six, thus being too weak below 50 GHz
to be observed. 

Using the frequencies here derived, we performed a search for C-cyanomethanimine emission toward nearby Sun-like-star forming regions using
the ASAI IRAM 30-m dataset. We investigated the earliest stages of the star forming
process, from starless regions to the more evolved hot-corinos
associated with both Class 0 and Class I objects.
We sampled spatial scales between 1600 and 2800 AU, depending
on the targets, obtaining the following upper limits 
on the C-cyanomethanimine column
density: $\sim$ 10$^{11}$ cm$^{-2}$ for the starless core L1544,
and fews 10$^{12}$ cm$^{-2}$ for the hydrostatic core Barnard 1, hot-cores, 
and shocked regions.   
Using previous knowledge on the H$_2$ column densities of the
observed regions, we derived C-cyanomethanimine abundances less than
a few 10$^{-10}$ and 10$^{-9}$ for starless and hot-corinos, and shocks, respectively. 
Finally, the present constraints on the C-cyanomethanimine abundance
could be used as instructive limits on the abundances in the gas-phase
of prebiotic N-bearing molecules more complex than HNCHCN, such as glycine.

\begin{acknowledgements}
This work has been supported in Bologna by MIUR `PRIN 2015' funds (project ``STARS in the CAOS (Simulation Tools for Astrochemical 
Reactivity and Spectroscopy in the Cyberinfrastructure for Astrochemical Organic Species)'' - Grant Number 2015F59J3R) and by the University 
of Bologna (RFO funds). This work has also been partially supported by the PRIN-INAF 2016  
"The Cradle of Life - GENESIS-SKA (General Conditions in Early Planetary Systems
for the rise of life with SKA)". The support of the COST CMTS-Actions CM1405 (MOLIM: MOLecules In Motion)
and CM1401 (Our Astro-Chemical History) is acknowledged. CC acknowledges the funding from the European Research Council (ERC), 
project DOC (contract 741002). JC thanks ERC for funding under ERC-2013-SyG, G.A. 610256 NANOCOSMOS. 
\end{acknowledgements}



\bibliographystyle{aa}
\bibliography{cyanomethanimineAA_Mattia}

\end{document}